\documentstyle[preprint,aps]{revtex}
\begin{document}
\draft
\tighten
\title{Calculation of exciton densities in SMMC}
\author{D.J. Dean$^{1}$ and S.E. Koonin$^2$}
\address{$^1$Physics Division, Oak Ridge National Laboratory\\ P.O. Box
2008, Oak Ridge, Tennessee 37831-6373 USA\\ and\\
Department of Physics and Astronomy, University of Tennessee\\ Knoxville,
Tennessee, 37996 USA\\
$^2$W. K. Kellogg Radiation Laboratory, California
Institute of Technology,\\ Pasadena, California 91125, USA}

\date{\today}
\maketitle

\begin{abstract}
We develop a shell-model Monte Carlo (SMMC) method to calculate densities
of states  with varying exciton (particle-hole) number.
We then apply this method
to the doubly closed-shell nucleus $^{40}$Ca in a full
$0s$-$1d$-$0f$-$1p$ shell-model space and compare our results to those
found using approximate analytic expressions for the partial densities.
We find that the effective
one-body level density is reduced by approximately 22\%
when a residual two-body interaction is included in the shell
model calculation.
\end{abstract}


\narrowtext

\begin{center}
{\bf 1. Introduction}
\end{center}
%\section{Introduction}

Particle-hole, or exciton, level densities enter into
the description of partial decay rates in nuclear
preequilibrium emission\cite{griffin66,Strohmaier97,FKK}. These level
densities have been modeled using analytic
expressions\cite{blann68,williams71}
that describe nuclear excitations in terms of the
number of particles, $p$, and holes, $h$, measured from the
Fermi surface, with the
exciton number $N_{e}=(p+h)/2$.

For a single species of particles, Williams \cite{williams71} derived an
expression for the partial density of states given by
\begin{equation}
\rho_{N_{e}}(E)=g\frac{\left(gE-G\right)^{2N_{e}-1}}
{p!h!\left(2N_{e}-1\right)!}\;,
\label{willf}
\end{equation}
where $E$ is the excitation energy measured above the
ground-state configuration, $g$ is the single-particle density of states,
and $G/g$ plays the role of an effective Pauli energy with
$G=(p^2+h^2)/4+(p-h)/4+h/2$.
There exist more complicated expressions that
distinguish between protons (with single-particle density
$g_p$) and neutrons ($g_n$)
in a given nucleus, but
we will not quote them here. In the most naive picture, a uniform
spacing of single-particle states, $d$, is assumed, in which case the
single-particle level density is $g=1/d$, measured in units of
MeV$^{-1}$.

Equation \ref{willf} and its neutron/proton
counterpart suffer from several deficiencies
including the assumption of an unlimited number of single-particle
states, an inexact treatment of the Pauli principle,
and the assumption of a uniform
single-particle level spacing. Extensions to the basic model that
ameliorate some of these effects have also been pursued. As examples,
we mention Bogilia et al. \cite{Bogilia96} in which the energy
dependence  of the single-particle level spacing was included in a
general way; using the equidistant single-particle picture,
Kalbach \cite{Kalbach89} and Zhang and Yang \cite{ZY88}
considered Pauli principle corrections to the state densities;
and De and Hua \cite{De93} considered the effects of pairing in addition
to the Pauli blocking on the state densities.
These effects were combined and extended to
non-uniform level spacings
by, for example, Harangozo et al. \cite{harangozo98}.

A further difficulty with the
simple formula is that the residual two-body interaction, which is
present beyond the nuclear mean field, and which includes important
contributions beyond $J=0$ pairing, is not incorporated. While
some progress has been made to approximate the effects of the
residual interaction \cite{pluhaf88} on the partial level densities,
no interacting shell-model calculations have been performed in
large model spaces that would indicate the effect
of the two-body interaction on the single-body density parameter, $g$,
nor have there been any partial density-of-state calculations in the
interacting shell model.

In this article, we describe calculations that study the effects
of the residual two-body interaction
on the partial level densities, and present results for
partial level densities in $^{40}$Ca. Our approach is
to study a related quantity, $Y_{N_{e}}(\beta)$,  which is
the ratio of the particle-hole partition functions, $Z_{N_{e}}(\beta)$,
to  the full partition function, $Z_A(\beta)$, as a function of the
inverse  temperature, $\beta$, (measured in MeV$^{-1}$) in the system.
We perform our calculations
in a full $0s$-$1d$-$0f$-$1p$ model space using shell-model Monte
Carlo (SMMC) techniques \cite{Koonin97,Lang93}, and an interaction
that describes reasonably well the low-lying spectral properties of
nuclei  in the $sd$-$fp$ region\cite{Dean98}. We will compare our results
with those obtained from eq.~(\ref{willf}) and its proton/neutron
counterparts. Finally, we will show partial densities of
states for several exciton numbers.
In section 2, we give an overview of our calculational method. We present
results in section 3, and conclude with a brief summary in section 4.

\begin{center}
{\bf 2.  Calculation of Excitons in SMMC}
\end{center}
%\section{Calculation of Excitons in SMMC}

Investigations into both ground-state and thermal properties of
nuclei have been described using the SMMC
technique \cite{Koonin97}.
This method offers an alternative description of
nuclear structure properties in
the shell-model context that is complementary to direct diagonalization.
SMMC is designed to give thermal or ground-state
expectation values for various
one- and two-body observables. Indeed, for larger nuclei, SMMC may be
the only way to obtain information
on the thermal properties of the system from a shell-model perspective.
In this method, we make use of
the imaginary time
many-body propagator $\hat{U}=\exp(-\beta\hat{H})$
to calculate the expectation values. For example,
the excitation energy of a nucleus is
$E(\beta)=\langle\hat{H}(\beta)\rangle-\langle \hat{H}(\infty)\rangle$,
where $\langle\hat{H}(\infty)\rangle$ is the ground-state energy.
In order to find the excitation energy of a nucleus with
particle number $A$, we must then calculate
\begin{equation}
\langle\hat{H}\rangle=
{{\hbox{Tr}\hat{P}_A\hat{U}\hat{H}}
\over{\hbox{Tr}\hat{P}_A\hat{U}}}\;,
{{\hbox{Tr}_A\hat{U}\hat{H}}
\over{\hbox{Tr}_A\hat{U}}}\;,
\end{equation}
where $\hat{P}_A=\delta(\hat{N}-A)$
projects the trace over all many-body states in the system to those
states that have the desired particle number.

Two-body terms
in $\hat{H}$ are linearized through the Hubbard-Stratonovich
transformation, which introduces auxiliary fields over which
one must integrate to obtain physical answers.
Since $\hat{H}$ contains many two-body terms that do not
commute, one must discretize $\beta=N_t\Delta\beta$.
The method can be summarized as
\begin{eqnarray}
Z_A&=&\hbox{Tr}_A\hat{U}=\hbox{Tr}_A\exp(-\beta\hat{H})
\rightarrow
\hbox{Tr}_A
\left[\exp(-\Delta\beta\hat{H})\right]^{N_t} \nonumber \\
&\rightarrow&
\int{\cal D}[\sigma]G(\sigma)\hbox{Tr}_A\prod_{n=1}^{N_t}\exp
\left[\Delta\beta\hat{h}(\sigma_n)\right]\; ,
\end{eqnarray}
where $\sigma_n$ are the auxiliary fields at a given imaginary
time-step $\Delta\beta$ (there is one $\sigma$-field
for each two-body matrix-element in $\hat{H}$ when the two-body terms are
recast in quadratic form),
${\cal D}[\sigma]$ is the measure of the integrand,
$G(\sigma)$ is a Gaussian in $\sigma$, and $\hat{h}$ is a one-body
Hamiltonian.  Thus, the shell-model problem is transformed
from the diagonalization of
a large matrix to one of large dimensional quadrature. Dimensions of the
integral can reach up to 10$^5$ for systems of interest, and it is thus
natural to use Metropolis
random walk methods to sample the space. Such integration is most
efficiently performed on massively parallel computers.
Further details are discussed in Koonin et al. \cite{Koonin97}.

In order to obtain density-of-state
information, we calculate in SMMC the expectation of the
energy and integrate the
thermodynamic relationship
\begin{equation}
E(\beta)=-\frac{d\ln Z_A(\beta)}{d\beta}\;
\end{equation}
to obtain
\begin{equation}
\ln Z_A(\beta)=-\int_{0}^{\beta}d\beta'E(\beta') - \ln Z_A(0)\;,
\label{full_z}
\end{equation}
where $Z_A(0)=\hbox{Tr}_A \bf{1}$ is the total number of $A$-particle
states in the system.  $Z_A(\beta)$ and $\rho(E)$ are related
by the inverse Laplace transform
\begin{equation}
Z_A(\beta)=\int_{-\infty}^{\infty} dE \exp(-\beta E) \rho(E)\;,
\label{laplace}
\end{equation}
which can be solved in a saddle point approximation to yield
\begin{eqnarray}
\rho (E)&=&\frac{\exp(S)}{\sqrt{2\pi\beta^{-2}C}} \nonumber \\
S&=&\beta E+\ln Z(\beta);\;\;\;\;\;\;\beta^{-2}C=-\frac{dE}{d\beta}\;.
\label{saddle_pt}
\end{eqnarray}
In this expression $C$ is the heat capacity of the system.

For our discussion, we study $^{40}$Ca.
The non-interacting
ground state is a filled $sd$-shell with
no particles in the $fp$-shell. Our excitons are then
enumerated with respect to the filled $sd$-shell.
We may excite both protons ($\pi$) and neutrons ($\nu$)
so that $N_{e}=(p_\pi+h_\pi+p_\nu+h_\nu)/2=N_{fp}$, where
$N_{fp}$ ($N_{sd}$) gives the number of particles in the $fp$- ($sd$-) shell.
For example, $N_{e}=2$ includes the following particle-hole
excitations:
($0p_\pi 0h_\pi$, $2p_\nu 2h_\nu$)
($1p_\pi 1h_\pi$, $1p_\nu 1h_\nu$)
($2p_\pi 2h_\pi$, $0p_\nu 0h_\nu$).
Furthermore, $0\leq N_{e} \leq 24$
since, at most, 24 particles can be excited from the $sd$-shell into
the $fp$-shell.

The ratio of the partition
function for $N_{e}$ excitons to the full partition function for the
$A$-particle system may be found by introducing a second number
projection operator,
\begin{equation}
\hat{P}_{N_{e}}=\delta(N_{sd}-\hat{N}_{sd})\delta(N_{fp}-\hat{N}_{fp})\;,
\end{equation}
provided that $A=N_{sd}+N_{fp}$. In reality we
perform this projection for both protons and
neutrons simultaneously, but this
only complicates notation and will not be discussed here.
We calculate the ratio of partition functions as
\begin{equation}
Y_{N_{e}}(\beta)=\frac{Z_{N_{e}}(\beta)}
{Z_A(\beta)}=\frac{\hbox{Tr}\hat{P}_{N_{e}}\hat{U}}
{\hbox{Tr}\hat{P}_A \hat{U}}\;.
\end{equation}
Therefore, $\sum_{N_{e}} Y_{N_{e}}(\beta)=1$ which we use
as a convenient numerical check.
We also extract the energy of
the particle-hole excitations, $E_{N_{e}}$, as
\begin{equation}
E_{N_{e}}(\beta)=-\frac{d\ln Y_{N_{e}}(\beta)}{d\beta}+E(\beta).
\label{eph}
\end{equation}
We may now employ eq.\ (\ref{saddle_pt}) for the partial density
of states:
\begin{eqnarray}
\rho_{N_e} (E)&=&\frac{\exp(S_{N_e})}{\sqrt{2\pi\beta^{-2}C_{N_e}}}
\nonumber \\
S_{N_e}&=&\beta_{N_e} E_{N_e}+\ln Z_{N_e}(\beta);\;\;\;\;\;\;
\beta_{N_e}^{-2}C_{N_e}=-\frac{dE_{N_e}}{d\beta}\;.
\label{sp_2}
\end{eqnarray}
Here $\beta_{N_e}=\beta_{N_e}(E_{N_e})$ is determined by inverting the relation
$E_{N_e}=E_{N_e}(\beta_{N_e})$, and $C_{N_e}$ is the heat capacity for the
particular exciton number.

\newpage

%\section{Results}
\begin{center}
{\bf 3.  Results}
\end{center}

We now turn to a description of our $^{40}$Ca calculation
in the $0s$-$1d$-$0f$-$1p$ shell model space.
Our starting point for an appropriate interaction is taken from
ref.~\cite{Dean98}. In order to obtain a microscopic effective
interaction, we begin with a free nucleon-nucleon interaction
which is appropriate for a description of low-energy nuclear structure.
The choice made in ref.~\cite{Dean98} was to work
with the charge-dependent version of the Bonn nucleon-nucleon
potential model as found in
ref.~\cite{cdbonn}.  Standard perturbation techniques, as discussed in
ref.~\cite{mhj}, were employed to obtain an effective
interaction in the full $sd$-$fp$ model space. Finally, the interaction was
modified in the monopole terms using techniques developed by
Zuker and co-workers \cite{az94,dz98}.

SMMC calculations for realistic
interactions typically have a Monte Carlo sign problem which
can be overcome by an extrapolation technique discussed
in ref.~\cite{alhassid94}, and successfully applied to the
$sd$-$fp$ region in \cite{Dean98}.
This extrapolation technique was also applied to
thermal properties of nuclei
\cite{dean95}, but the statistical error inherent
in the energy upon extrapolation
prevents a full description of the density of states unless one
has good justification to spend the computational resources to
reduce the statistical error.
It was recently demonstrated that a good reproduction of the experimental
density of states could be obtained
for nuclei in the $0f1p$-$0g_{9/2}$ shell \cite{alhassid97} using a
pairing-plus-quadrupole interaction that was free from the
sign problem. In this work, we fit our realistic two-body interaction
discussed above to a pairing-plus-multipole interaction given by
\begin{eqnarray}\label{1}
\hat{H}_2 = g_0 \pi \hat{P}^\dagger_{00}\hat{P}_{00}
+4\pi \sum_{\nu\mu} \chi_{\nu}
:  \sum_\mu (-)^{\mu} \hat{Q}_{\nu\mu}  \hat{Q}_{\nu -\mu}: \;,
\end{eqnarray}
where $::$ denotes normal ordering and
$\hat{P}^\dagger_{\lambda \mu}, \; \hat{Q}_{\nu\mu}$
are pair and quadrupole operators given by
\begin{eqnarray}\label{2}
\hat{P}^\dagger_{\lambda \mu} & = &
\sum_{a b} (-)^{\ell_b} (j_a \parallel { \cal Y_\nu} \parallel j_b)
[ \hat{a}_{j_a}^{\dagger} \times \hat{a}_{j_b}^{\dagger} ]_{\lambda \mu}
\nonumber \\
\hat{Q}_{\nu\mu} & = & -{1\over \sqrt{2\nu+1}} \sum_{a c} (j_a \parallel
r^\nu {\cal Y}_{\nu} \parallel j_c)
{[\hat{a}^{\dagger}_{j_a }\times \hat{\tilde{a}}_{j_c}]}_{\nu\mu} \;.
\end{eqnarray}
In eq.\ (\ref{2}) $a\equiv n \ell j$ denotes a single-particle orbit
and $\hat{\tilde{a}}_{j m} = (-)^{j+m} \hat{a}_{j -m}$.
We fit $g_0$ and the $\mu=2,4,6$ multipoles to the realistic
interaction from \cite{Dean98}. A least-squares fit gives an interaction
which indeed has a good Monte Carlo sign. After some minor
adjustments to the pairing
strength in order to obtain a better gap between ground states and first
excited states in several light nuclei, we use the following parameter
set: $g_0=-0.63$~MeV, $\chi_2=-0.047$~MeV fm$^{-4}$,
$\chi_4=-0.001$~MeV fm$^{-8}$, and $\chi_6=-0.17\times10^{-3}$~MeV fm$^{-12}$.
(Large enhancements of the $\langle a \mid r^\nu \mid b \rangle$ matrix
elements as $\nu$ increases is the reason for the decrease in $\chi_\nu$
values, although contributions to two-body matrix elements arising
from the higher multipoles is significant.)
Our single-particle
energies are 0.0, 5.36, 0.64, 8.21, 14.21, 10.14, and 12.07~MeV for the
$0d_{5/2}$, $0d_{3/2}$, $1s_{1/2}$,
$0f_{7/2}$, $0f_{5/2}$, $1p_{3/2}$, and $1p_{1/2}$ orbitals, respectively.
We do not correct for center-of-mass motion in these calculations,
although such a contamination to the $Y_{N_e}$ should be fairly small in
this system. Furthermore, we do not include odd multipoles which upon
fitting were found to give coefficients that cause Monte Carlo sign
problems. Thus our negative parity states are probably less well
described by this  choice of Hamiltonian.

We use the results of the non-interacting case to demonstrate the
validity of our technique for finding the partial partition functions.
In order to show this,
we calculate by enumeration the total number of many-body states
for each $N_{e}$. This can most easily be done
by using eq.~(\ref{full_z}) to find $Z_A(\beta=0)$.
We also find $Y_{N_{e}}(0)$ by an extrapolation from
small, but finite, $\beta$. We show in fig.~\ref{fig1} our results
for the number of states as a function of the
exciton number.
The SMMC results are compared to an exact counting of the number
of states of a given exciton number. The agreement is excellent.
The total number of calculated SMMC states is
3.834$\times$10$^{16}$ as compared to the
exact value of 5.095$\times$10$^{16}$.

We show in fig.~\ref{fig2} a comparison of the non-interacting (left)
and interacting (right) calculation.
The $N_{e}=0$ calculation gives
some indication of the thermal freeze-out of the ground state.
The non-interacting calculation requires fairly large
$\beta$ to fully reach the ground state, since the first
excited state is only 0.64~MeV above the ground state. We pursued
these calculations to $\beta=4.0$~MeV$^{-1}$, for which
$\langle H\rangle$ is 0.139~MeV from the ground-state value.
Since it takes more thermal energy to overcome the pairing interaction
and to excite them from the
ground-state configuration, the interacting
curves representing different $N_{e}$ excitations
are compressed in $\beta$ relative to the non-interacting curves.  (The
excitation energy of the first excited state is
approximately 3.5~MeV.)  However,
the same features remain. Clearly, as thermal energy is decreased,
and $\beta$ becomes larger (temperature decreases),
it is more difficult to produce large
particle-hole excitations in the system.
The converse is also true at higher
temperatures, where it is difficult to
obtain only $N_{e}=2$, for example.
In the interacting case, the low temperature tail of the
$N_e=2$ exciton tends to spread further in $\beta$ than
does the $N_e=1$ tail.  As we shall see, this has
direct consequences on the partial densities of states.
Furthermore, since
$\sum_{N_{e}} Y_{N_{e}}(\beta)=1$, we can interpret $Y_{N_{e}}$ as a measure
of likelihood to find $N_{e}$ excitons at
a given temperature. Since the excitation energy is a
monotonic function of the temperature, one expects the density of states to
be dominated by excitons of a particular type in a given energy range.
As we shall see, this is indeed the case.

We compare our results to those obtained from eq.~(\ref{willf}) and
its proton/neutron equivalent by adjusting $g$ to obtain a fit
to the calculated $Y_{N_e}$ curves. This is demonstrated for the
$N_e=4$ curve in fig.~\ref{fig3}.
We fit each of our SMMC curves to eq.~(\ref{willf}) for
an effective one-body density parameter $g_{\rm eff}=g_p+g_n$
\cite{ms69}, and
the effective level density parameter is $a=(\pi^2/6)g_{\rm eff}$.
A uniform Fermi gas yields
$a\approx A/15$
(=2.67 for A=40)~MeV$^{-1}$, a harmonic oscillator potential yields
$a\approx A/10$ MeV$^{-1}$ (=4.0), and the empirical value is
$A/8$~MeV$^{-1}$.
We obtain $a=5.43$ ($g_{\rm eff}=3.3$) MeV$^{-1}$
for the non-interacting case, and
$a=4.27$ ($g_{\rm eff}=2.6$) MeV$^{-1}$ in the interacting
case, rather independent (within 0.02 MeV$^{-1}$)
of $N_e \ge 2$. For $N_e=1$ the comparison cannot
be made as the Blann-Williams formula breaks down.
Thus, $g_{\rm eff}$ is reduced by 22\% in the presence of
an interaction.

The decomposition of the $N_e=4$ case into the various proton-neutron
components is also shown in fig.~\ref{fig3}. The
($0p_\pi 0h_\pi$, $4p_\nu 4h_\nu$) plus its neutron counterpart carries very
little weight here, while the ($2p_\pi 2h_\pi$, $2p_\nu 2h_\nu$) component of
$Y_4$ carries the most weight. As expected, in all
$N_e$ cases the largest component
of $Y_{N_e}$ is the one in which the number of excited neutrons equals the
number of excited protons.

Finally we present our result for the
calculation of $\rho_{N_{e}}$
derived by using $Z_{N_{e}}(\beta)$ in eq.~(\ref{full_z}) and
$E_{N_{e}}$ from eq.~(\ref{eph}).
The natural log of $\rho_{N_{e}}$
is shown in fig.~\ref{fig4} for the $N_{e}$=1--6 excitations.
We also include in the figure the total state density as a function of
the excitation energy in the system. The saddle point approximation
breaks down in regions where there are very few states, which
makes it difficult to describe well-separated states in
the low-lying spectrum ($E^*<3$~MeV) for
the full density or for the individual exciton densities.
We also propagated our statistical error bars through the calculation
of $\rho_{N_e}$, but, as can be seen, they are very small except in the
case of the $N_e=1$ excitons. Note that
the majority of states, for example at $E^*=25$~MeV, are $N_{e}=4$ states,
while at $E^*=35$~MeV the $N_{e}=5$ states contribute most.
This localization in excitation energy
of excitons was reflected in our earlier discussion
of the behavior of $Y_{N_{e}}$. Interestingly, the $N_e=1$ density of states
begins in energy slightly above 2p2h state
density. Recall that experimentally the first
excited state of $^{40}$Ca is a 0$^+$ 2p2h state (at 3.3 MeV), and that the
first negative parity state (a 3$^{-}$)
occurs at a slightly higher energy of 3.7 MeV.  Our Hamiltonian
fairly closely gives the correct relative starting energies for these two
exciton configurations, although, due to the breakdown
of the saddle-point approximation for low state densities,
we cannot precisely determine the excitation energy of the first excited
0$^{+}$ level.  We also note an
interesting pairing effect that shows up in the partial densities.
Note that the $N_e=1$ state density starts about 0.6~MeV above the
$N_e=2$ case. The $N_e=3$ state density begins approximately 2~MeV above
the $N_e=4$ case. This is a manifestation of pairing in the system. It
takes more excitation energy to produce an odd particle-hole excitation
than it does to produce an even particle-hole excitation since energy must
be  expended to break pairs for odd excitations. This effect was already
apparent from the discussion of the low-temperature behavior of the
$Y_{N_e}$, as indicated above.

%\section{Conclusion}
\begin{center}
{\bf 4.  Conclusion}
\end{center}

We have discussed in this article how one may obtain information
on particle-hole excitations using SMMC methods
for calculations of a nuclear system. Our technique uses
the ratio of the particle-hole partition function to the full partition
function of the system.
The method incorporates exact Pauli blocking,
non-equidistant single-particle energies, and gives the exact
partial densities for a given nuclear effective interaction, within
statistical errors. It also
has a well-defined energy scale. One drawback of the present calculation
is that the space size is limited to two major oscillator shells, although
this can be rather easily overcome.
The projection operator introduced in this may be applied
in any Monte Carlo technique where ratios of partition functions
are needed. Our results also indicate that the effective $g$-parameter used
in eq.\ (1) should be reduced by approximately 22\% to
account for the inclusion of the two-body interaction which
acts to correlate the nucleus beyond the simple mean-field or pairing
prescription.

The method we have described here could be further advanced in two
ways. One may increase the model space used, thus allowing
for a broader range of energies and excitation modes to be explored.
The method is also applicable to $mp$-$nh$ excitations if we extend
our studies to open-shell nuclei such as, e.g., $^{42}$Ca. This may
be pursued in future work.

\acknowledgements

This work was supported in part through grant DE-FG02-96ER40963
from the U.S. Department of Energy.
Oak Ridge National Laboratory (ORNL) is managed by Lockheed Martin Energy
Research Corp.\ for the U.S. Department of Energy under contract number
DE-AC05-96OR22464.  We also acknowledge support from the
U.S. National Science Foundation
under Grants No. PHY-9722428, PHY94-12818 and PHY94-20470.

\begin{figure}
\caption{Top panel: Running sum of the non-interacting states found
from $Z_{N_{e}}$
at $\beta=0$ compared with the
running sum from an exact calculation. Bottom
panel: the total number of states for each
$N_{e}$ found in both non-interacting
SMMC calculation and compared to the exact result.
}
\label{fig1}
\end{figure}

\begin{figure}
\caption{A comparison of the interacting (right)
and non-interacting (left)
functions $Y_{N_{e}}$ for $N_{e}=0,\cdots,6$.
}
\label{fig2}
\end{figure}

\begin{figure}
\caption{Decomposition of the $Y_{N_e=4}$ into its particle hole components,
and a comparison of the full $Y_{N_e=4}$ generated from SMMC
to that obtained from eq.~(\protect\ref{willf}). The fit corresponds to
$g_{\rm eff}=2.6$~MeV$^{-1}$.
}
\label{fig3}
\end{figure}

\begin{figure}
\caption{Calculated partial densities of states $\rho_{N_e}(E_{N_e})$
for $N_{e}=1,\cdots,6$, using
the saddle point approximation from eq.~(\protect\ref{sp_2}).
Also plotted is the total state density calculated from
eq.~(\protect\ref{saddle_pt}). Statistical error bars that are
not visible are smaller than the symbols used.
}
\label{fig4}
\end{figure}

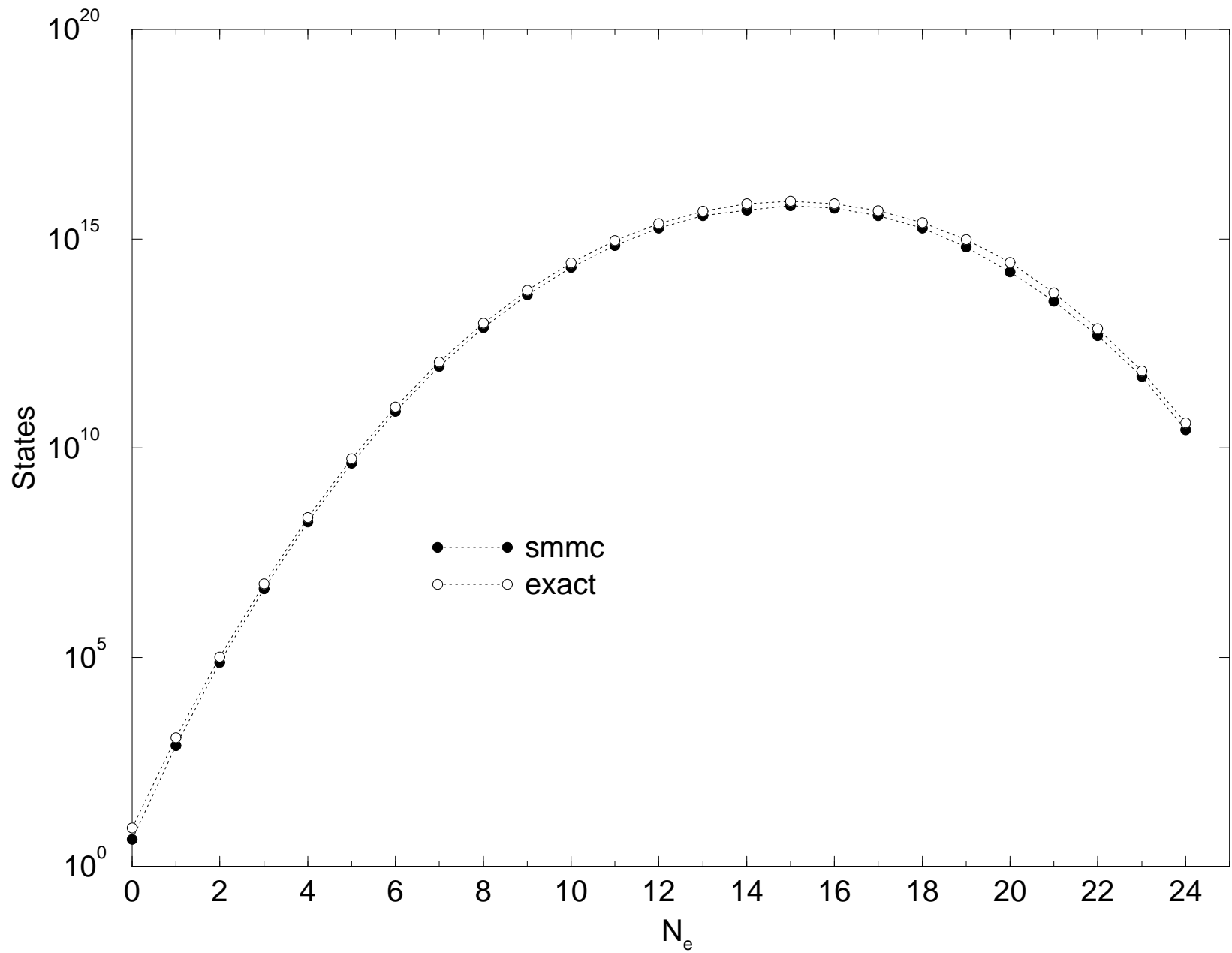
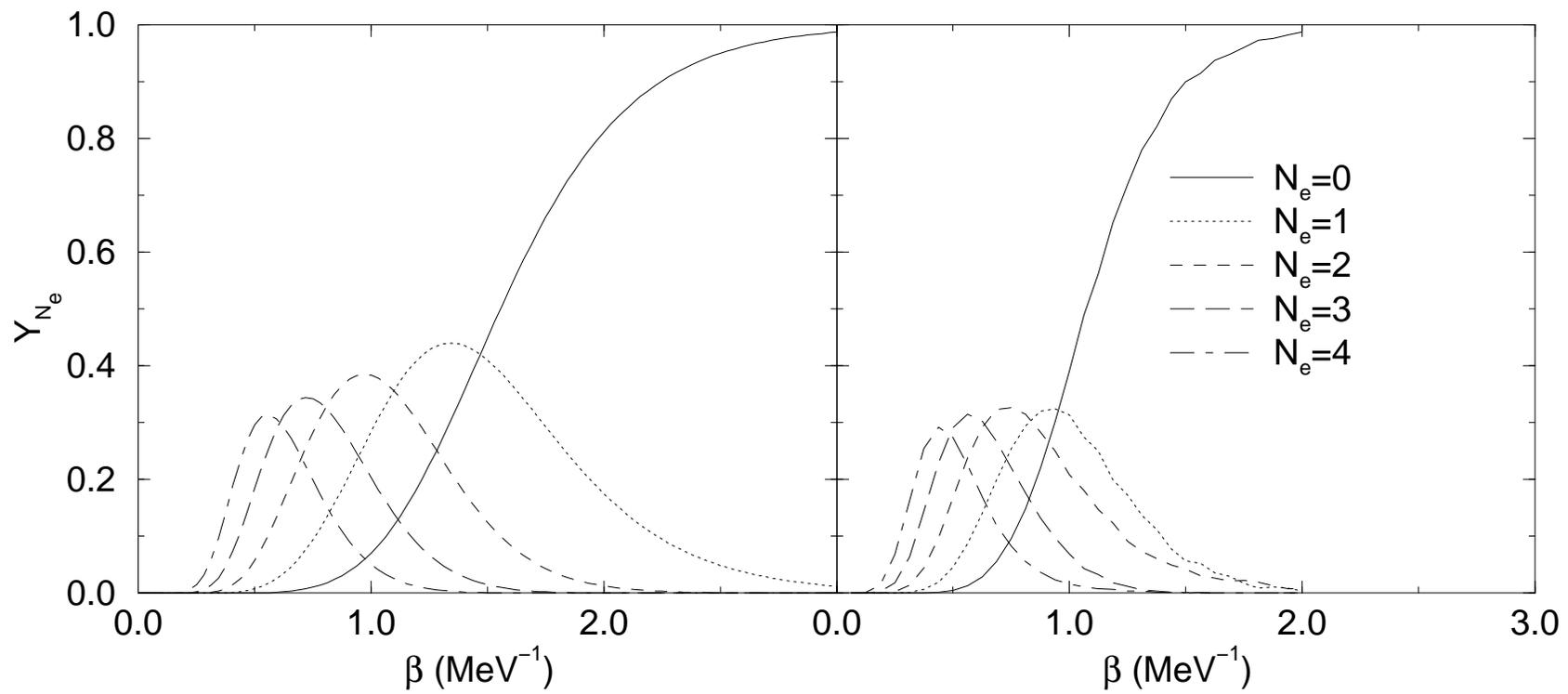
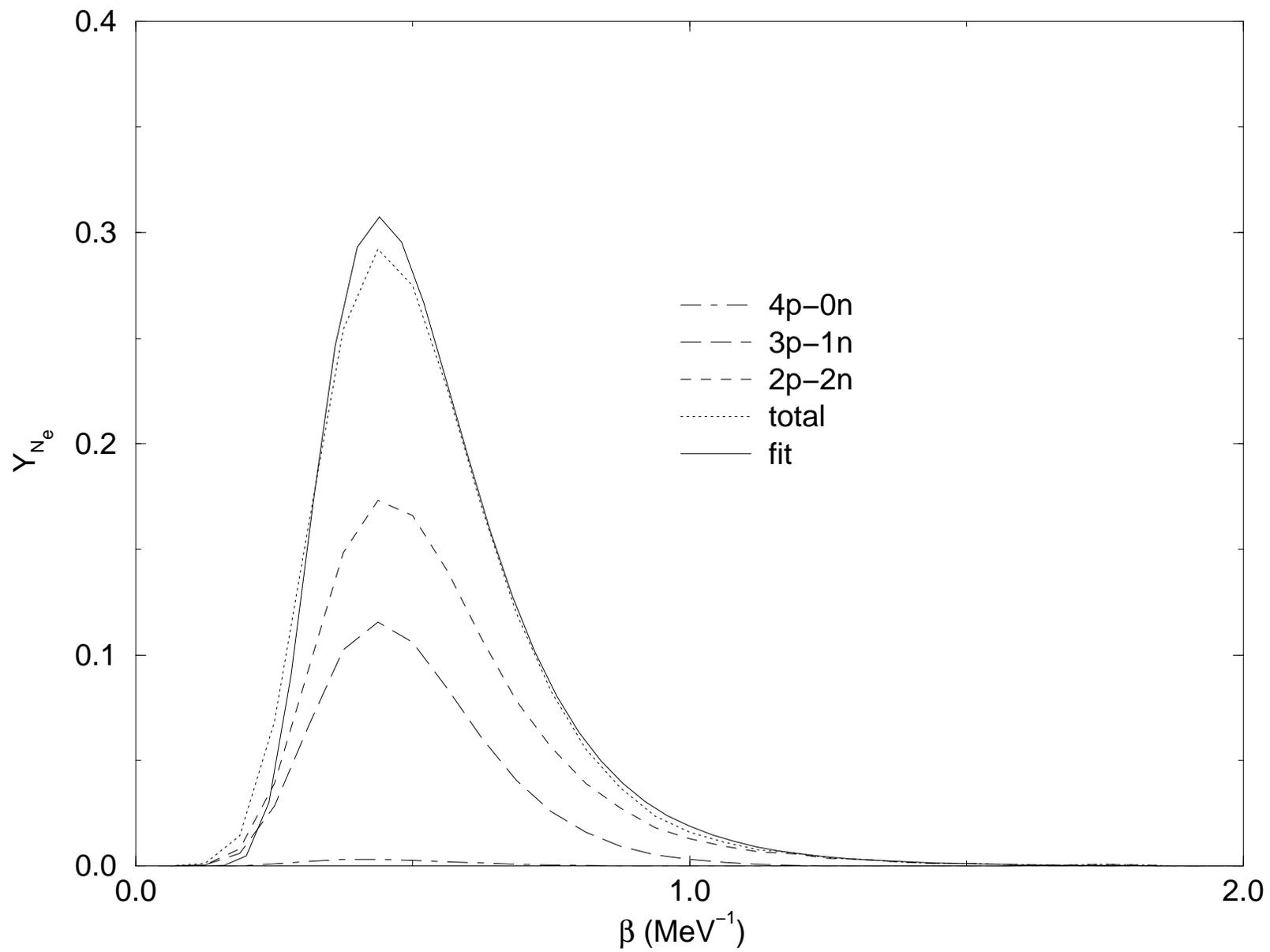
\end{document}